# Switching on and off the spin polarization of the conduction band in antiferromagnetic bilayer transistors

Fengrui Yao[1,2,#,*], Menghan Liao[1,2,#], Marco Gibertini[3,4*], Cheol-Yeon Cheon[1,2], Xiaohanwen Lin[1,2], Fan Wu[1,2], Kenji Watanabe[5], Takashi Taniguchi[6], Ignacio Gutiérrez-Lezama[1,2] and Alberto F. Morpurgo[1,2*]

[1]*Department of Quantum Matter Physics, University of Geneva, 24 Quai Ernest Ansermet, CH-1211 Geneva, Switzerland*

[2]*Group of Applied Physics, University of Geneva, 24 Quai Ernest Ansermet, CH-1211 Geneva, Switzerland*

[3]*Dipartimento di Scienze Fisiche, Informatiche e Matematiche, University of Modena and Reggio Emilia, IT-41125, Modena, Italy*

[4]*Centro S3, CNR-Istituto Nanoscienze, IT-41125, Modena, Italy*

[5]*Research Center for Electronic and Optical Materials, National Institute for Materials Science, 1-1 Namiki, Tsukuba, 305-0044, Japan*

[6]*Research Center for Materials Nanoarchitectonics, National Institute for Materials Science, 1-1 Namiki, Tsukuba, 305-0044, Japan*

[#] these authors contribute equally

*Correspondence: fengrui.yao@unige.ch; marco.gibertini@unimore.it; alberto.morpurgo@unige.ch

**Antiferromagnetic conductors with suitably broken spatial symmetries host spin-polarized bands, which lead to transport phenomena commonly observed in metallic ferromagnets. In bulk materials, it is the given crystalline structure that determines whether symmetries are broken and spin-polarized bands are present. Here we demonstrate experimentally that double-gate transistors realized on bilayers of van der Waals antiferromagnetic semiconductor $CrPS_4$ allow the relevant symmetry to be controlled by a perpendicular electric displacement field. Such a level of control enables the spin-polarization of the conduction band to be switched on and off. Because conduction band states with opposite spin-polarizations are hosted in the different layers and are spatially separated, these devices also give control over the magnetization of the electrons that are accumulated electrostatically. Our experiments show that double-gated $CrPS_4$ transistors provide a viable platform to create gate-induced conductors with near unity spin polarization at the Fermi level, as well as devices with a full electrostatic control of the total magnetization of the system.**

In antiferromagnetic conductors, spin order breaks time-reversal ($\hat{T}$) symmetry. However, if a time reversal symmetry transformation followed by spatial inversion ($\hat{P}$) or by a translation remains a symmetry (so-called crystal time-reversal symmetry) antiferromagnets can behave as if $\hat{T}$ symmetry was effectively present[1-5]. These considerations are key for antiferromagnetic spintronics[6-9], since the breaking of such "effective" time reversal symmetry causes physical phenomena characteristic of ferromagnets (anomalous Hall effect[10-13], spin-polarized bands[14-17], etc.), despite the absence of a net magnetization. Identifying bulk antiferromagnets that exhibit these phenomena requires a detailed analysis of the crystalline and magnetic structures, to determine whether $\hat{T}$ symmetry is effectively broken[2, 10]. Symmetry considerations analogous to the ones just mentioned are leading to unanticipated results, such as the discovery of altermagnetic compounds[3, 4, 13, 16, 17], and are responsible for the rapid development of antiferromagnetic spintronics.

It has been proposed that in some two dimensional (2D) antiferromagnetic semiconductors[18-23], spatial symmetries can be controlled experimentally enabling switching on and off at will the effect of time-reversal symmetry[24-28]. This is the case for bilayers of A-type antiferromagnetic semiconductors, in which the magnetization of one layer is equal and opposite to that of the other layer. Their low-energy conduction band is formed by spin-degenerate states residing in either one of the two layers, whose spin-direction is determined by the magnetization of the corresponding layer. The conduction band therefore consists of spin-unpolarized bands, associated to spatially separated electronic states. Under a perpendicular electric displacement field ($D$), inversion and space-time symmetry are broken, as the interlayer electrostatic potential difference shifts states with one spin to energies lower than states with the opposite spin (Fig. 1a and Supplementary Note 1). The conduction band becomes spin-polarized, and the spin-polarization can be reversed by inverting $D$.

The spatial separation of spin-split bands also gives control over their electronic population, by using double-gate field effect transistors (FETs) that allow tuning independently $D$ and the accumulated charge density ($n_e$)[29-33]. At zero electric field ($D/\varepsilon_0 = 0$), the potential of the two layers is the same, so that spin up and down bands are equally populated (see Supplementary Fig.1 and Supplementary Note 1). When the two gate electrodes are biased asymmetrically, the bands in the two layers shift to different energies. At low $n_e$, the added electrons go to the layer hosting the lowest conduction band edge and are fully spin polarized. Here, we investigate experimentally double-gated FETs based on $CrPS_4$ antiferromagnetic bilayers (see Fig.1b and Extended Data Fig. 1 for the schematics of such a device) to controllably generate and populate gate-tunable spin-split bands, and to detect them by measurements of hysteretic magnetoconductance.

**Double-gated bilayer $CrPS_4$ transistors**

$CrPS_4$ is a weakly anisotropic A-type vdW semiconducting antiferromagnet with Néel temperature $T_N = 38$ K[34-39]. Bulk crystals exhibit spin-flip and spin-flop transitions respectively at 7-8 T and 0.6-0.8 T (exact values depend on the crystals investigated). In thick $CrPS_4$ multilayers single-gate FETs, magnetoconductance measurements[38, 39] have allowed detecting the influence of the magnetic state on the band structure. In bilayers, electronic structure calculations (see Supplementary Fig. 1) show that conduction band states with opposite spin are spatially separated (as needed to create spin-polarization by applying a perpendicular electric field). So far, however, no transport measurements have been reported on double-gate bilayer transistors of $CrPS_4$, or of any antiferromagnet (pioneering work on $CrI_3$ bilayers focused on magnetooptical studies[31-33], because the insulating behavior of $CrI_3$ prevented transport measurements[40]).

Fig. 1c, d show transfer curves of a double-gate bilayer $CrPS_4$ device (for thickness identification see Extended Data Fig. 2) measured as a function of voltage applied to top and bottom gate. Two-terminal measurements are performed employing exfoliated graphite strips as contacts. On thick multilayers, we succeeded in realizing multiterminal transistors showing that two- and four-terminal measurements give virtually identical results at large source-drain bias[41] (these multiterminal devices also show the absence of Hall effect, as it often happens in accumulation layers of low-mobility semiconductors[42]).

**Doping-dependent magnetism at zero electric displacement field**

Creating and probing spin-polarization is effectively achieved by operating double-gate transistors to have large perpendicular electric field $D/\varepsilon_0$ and small accumulated electron density $n_e$ (see Fig. 1a). A large $D/\varepsilon_0$ maximizes the inter-layer potential difference, responsible for the energy difference between opposite spin bands. A small $n_e$ allows populating states with only one spin direction, and ensures that the accumulated electrons do not affect the magnetic state[38].

Prior to exploring this regime, we characterize the devices at zero displacement field. The magnetoconductance $\delta G(H) = (G(H) - G_0)/G_0$ ($G(H)$ is the conductance measured at magnetic field $\mu_0 H$, and $G_0 = G(H = 0)$) of a $CrPS_4$ double gated FET measured at zero displacement field is shown in Fig. 2, with magnetic field applied either perpendicular (out of plane, Fig. 2a and Fig. 2b) or parallel (in plane, Fig. 2c and Fig. 2d) to the layers. At low $n_e$, the spin-flip field ($H_{flip}$) is approximately 3.5 T (the field at which the magnetoconductance starts to flatten), half the bulk value, because each constituent layer feels the exchange interaction of only one neighboring layer[43, 44]. The precise value is determined by the minimum in $dG^2/dH^2$ (see Supplementary Fig. 2 and Supplementary Fig. 3), from which we see that $H_{flip}$ decreases pronouncedly upon increasing $n_e$, and that $H_{flip}$ is larger when the field is applied in-plane (see Fig. 2e and Supplementary Fig. 2 for details). The flat magnetoconductance observed at low magnetic field (see Fig. 2b) also allows determining the spin-flop field $H_{flop}$ [38, 43] (see Fig. 2f).

From these measurements we extract quantitative values for the parameters determining the magnetic state of the $CrPS_4$ bilayer, the interlayer exchange energy $J$ and the uniaxial magnetic anisotropy $K$ (see Methods and Supplementary Note 2). To this end, we express the magnetic energy of the system as $E = J\mathbf{M}_1 \cdot \mathbf{M}_2 / M_s^2 - K/2 \, (M_{1z}/M_s)^2 - K/2 \, (M_{2z}/M_s)^2 - \mu_0 \mathbf{H} \cdot (\mathbf{M}_1 + \mathbf{M}_2)$, where $\mathbf{M}_1$ and $\mathbf{M}_2$ are the magnetizations of the two layers due to the Cr atoms, $M_s$ is the single-layer saturation magnetization (per unit cell) and $\mathbf{H}$ is the applied magnetic field.

The analysis of the spin-flip transition with in- and out-of-plane field (Fig. 2e) gives the values of $J$ and $K$ shown in Fig. 2g and Fig. 2h, both decreasing upon increasing $n_e$. Surprisingly, the magnetic anisotropy vanishes at $n_e > 7\text{-}8 \times 10^{12}$ cm$^{-2}$ ($K$ likely changes sign and the magnetization reorients to be in the plane[45], but under these conditions our magnetoconductance measurements cannot be used to determine its value). The observed trends agree with ab-initio calculations (Fig. 2g). The suppression of $J$ can be understood as due to the already established downshift of the conduction band edge in the ferromagnetic state[39]. Because of this downshift, accumulating electrons increases the energy of the antiferromagnetic state ($E_{AFM}$) more than that of the ferromagnetic one ($E_{FM}$) so that the interlayer exchange energy $J$ (related to $E_{FM}(n_e) - E_{AFM}(n_e)$) decreases.

We conclude that –when probing the existence of spin-polarized bands in CrPS$_4$ bilayer– the charge density needs to be limited to values well below $7 \times 10^{12}$ cm$^{-2}$, to avoid that the magnetic state of CrPS$_4$ is strongly affected by the accumulated electrons. We also conclude that the influence of the accumulated electrons on the magnetic state is well described by a large change in the values of $J$ and $K$, and not in the magnetization of the layer where electrons are accumulated. Indeed, as $n_e$ is increased up to approximately $10^{13}$ cm$^{-2}$, $J$ changes by a factor of 2 and $K$ vanishes, whereas the change in layer magnetization is smaller than 0.5% (the electron density is only approximately 1% of the density of Cr atoms and the spin of Cr is 3/2). We should therefore analyze the system by considering first that electrons populate states in the conduction band associated with the underlying magnetic structure created by the Cr atoms (determined considering that $J$ and $K$ are function of $n_e$), and only later consider the effect of the modified layer magnetization.

**Hysteretic magneto-transport**

In the presence of a perpendicular displacement field and sufficiently low $n_e$, electrons are accumulated in the bottom or top layer depending on the sign of $D$. They occupy one of the spin-split bands, whose spin polarization is determined by the magnetization of the Cr atoms in the same layer (see Fig. 1a). The system has then two possible states (labeled A and B) with opposite spin of the accumulated electrons (Fig. 3a). The two states are energetically degenerate at zero applied perpendicular magnetic field $H_\perp$, and their energies shift in opposite direction when sweeping $H_\perp$ towards positive or negative values, due to the Zeeman energy of accumulated electrons. Even if the system is initialized in the low-energy state, therefore, it will eventually occupy the high-energy metastable state when the magnetic field is swept and changes sign. More specifically, when the magnetic field is swept from large negative values through the (negative) spin-flop field[46], there is no preference between states A and B if $D/\varepsilon_0 = 0$. With a finite $D$, however, the system develops a preference, and favors the state with lower energy (as shown in Fig.3a). As the magnetic field is swept to positive values, the system switches to the other state, resulting in an antiparallel magnetization arrangement, but with the Néel vector reversed (see Extended Data Fig. 3 for details). We then expect the magnetoconductance to be hysteretic, if the two states with opposite electron magnetization exhibit different conductance.

Fig. 3b shows the low-field magnetoconductance measured in a perpendicular magnetic field, at $n_e = 1.5 \times 10^{12}$ cm$^{-2}$, for different values of $D/\varepsilon_0$. Hysteresis emerges at finite $D/\varepsilon_0$, so that the magnetoconductance differs depending on whether the applied magnetic field is swept from negative to positive (blue curves) or from positive to negative (red curves) values. The hysteresis ends with a sharp jump at approximately $\mu_0 H = 0.2$ T, exhibiting a phenomenology typical of easy axis ferromagnets[47]. No hysteresis is observed for parallel magnetic field, as expected, since in a parallel field states A and B always have the same energy (see Extended Data Fig. 4 for details). The observed behavior therefore confirms that electrons generate a net magnetization as they populate the spin-split conduction band of the CrPS$_4$ bilayer, owing to the symmetry breaking induced by the applied displacement field (virtually identical behavior has been seen in another double-gated device, see Supplementary Fig. 4). The switching between the two magnetic states may occur with the bilayer staying in a single domain (as in the Stoner-Wolfarth model[47]), or by breaking the CrPS$_4$ bilayers into magnetic domains[47]. In the latter case, reversing the magnetic field sweep direction halfway the hysteresis loop should result in "minority" hysteresis loops, with the magnetoconductance that does not re-trace the curve measured when the magnetic field is swept up to $H > H_{\text{flop}}$. This is indeed what we observe experimentally (see Supplementary Fig. 5).

Fig. 3c shows how the magnitude of the magnetoconductance hysteresis evolves upon increasing displacement field $D/\varepsilon_0$, with data measured at a fixed carrier density $n_e = 1.5 \times 10^{12}$ cm$^{-2}$. Starting from $D/\varepsilon_0 = 0$, $\delta G_\uparrow - \delta G_\downarrow$ first increases rapidly, before saturating at approximately $D/\varepsilon_0 = 0.18$ V/nm. The dependence of the magnitude (quantified by the peak value of $\delta G_\uparrow - \delta G_\downarrow$) is summarized by the brown dots in Fig. 3d. Measurements at larger values of accumulated electron density $n_e$ exhibit the overall same behavior (see the red and blue dots in Fig. 3d), but the displacement field needed to reach saturation increases. The trend is consistent with the behavior expected for a conduction-band spin-splitting induced by the displacement field (as shown in Fig. 1a). As $D/\varepsilon_0$ is initially turned on, the spin-splitting is small –much smaller than the Fermi energy corresponding to the accumulated charge density– so that both the spin up and the spin down bands in the two layers are populated. The population of the two bands is only slightly different, which is why the amplitude of the hysteresis is small. As $D/\varepsilon_0$ is increased, the spin-splitting in the conduction band also increases, and so does the difference in population of spin up and down bands, which is why the amplitude of $\delta G_\uparrow - \delta G_\downarrow$ also increases. At sufficiently large $D/\varepsilon_0$, the splitting between the spin up and down bands becomes larger than the Fermi energy, so that the electrons populate only one of the bands. Past this point, a further increase in $D/\varepsilon_0$ does not change the population of the spin-split bands, and $\delta G_\uparrow - \delta G_\downarrow$ saturates.

This scenario naturally explains why the accumulation of a larger electron density $n_e$ requires a larger displacement field to reach saturation. We estimate the value of displacement field at saturation as $D_{\text{sat}}/\varepsilon_0 = n_e \varepsilon_r/(e\, d\, (m*/2\pi\hbar^2))$, by equating the induced electrostatic energy difference between the two layers to the Fermi energy of the electrons occupying one layer ($d$ is the distance between the CrPS$_4$ layers forming the bilayer, $\varepsilon_r$ is the relative dielectric constant (3.9) and $m^*$ is the effective mass (1.26 times the free electron mass) in CrPS$_4$[48], $e$ is the electron charge and $\hbar$ Planck's constant). For $n_e = 1.5 \times 10^{12}$ cm$^{-2}$, $D_{sat}/\varepsilon_0 \approx 0.3$ V/nm, close to the experimental value. Both the argument invoked to explain the evolution of the magnetoconductance hysteresis with displacement field, and the good correspondence between the estimated and measured values of $D_{sat}/\varepsilon_0$, support the scenario that the conduction band is fully spin polarized for $D/\varepsilon_0 > D_{sat}/\varepsilon_0$.

**Odd-even effects**

To confirm that a vertical displacement field causes a hysteretic magnetoconductance because of the effect of inversion symmetry breaking on the accumulated electrons, we have explored devices realized on 3, 4, and 5 CrPS$_4$ layers (3 L, 4 L, and 5 L; see Fig. 4). Even (2 L and 4 L) and odd (3 L and 5 L) CrPS$_4$ multilayers should exhibit distinctly different behavior, because in odd layers the magnetization of the top and bottom layers is the same, whereas in even layers it is opposite. Reversing the displacement field polarity, therefore, alters the spin polarization of gate-accumulated electrons only in even layers. Additionally, in odd multilayers the magnetization of the uncompensated layer of Cr atoms ($M_{Cr} \neq 0$) is nearly thousand times larger than that of the accumulated electrons. Therefore, in odd multilayers the magnetization is large already in the absence of any accumulated electron, and switches at small applied magnetic field (around 0.05 T [37]). In other words, in odd layers, the switching between state A and state B (see schematics in Fig. 3 for bilayers) is driven by the uncompensated layer's magnetization: since the conductance does not depend on whether the Cr magnetization points up or down, the switch between state A and B has no measurable effect on transport.

Fig. 4 shows magnetoconductance data for 3-5 L CrPS$_4$. At $D/\varepsilon_0 = 0$, no hysteresis is observed below the spin-flop field irrespective of layer thickness, as expected. A very pronounced odd-even effect is evident in the spin-flop transition fields, with even layers transitioning at approximately 0.5 T and odd layers at 2-3 times larger field (Fig. 4a). This difference originates from the magnetization of uncompensated layer present in odd multilayers(Fig. 4b and Fig. 4c), which shifts the spin-flop transition to higher magnetic field (see earlier work on CrSBr and CrCl$_3$[43, 49]). More importantly, when an electric field is applied, the behavior of even and odd layers becomes markedly different. Hysteresis in the magnetoconductance emerges, but only in even layers (Fig. 4d): odd layers continue to show no hysteresis (Fig. 4e). The amplitude of the magnetoconductance hysteresis in 4 L CrPS$_4$ is comparable to that of 2 L CrPS$_4$, even though detailed aspects are different. For instance, increasing the displacement field from a negative (-0.6 V/nm) to a positive value (0.6 V/nm) causes the sign of the magnetoconductance hysteresis in 4 L to change multiple times (Supplementary Fig. 6). This is likely because the electronic wavefunctions evolve from being distributed equally over all layers at $D/\varepsilon_0 = 0$ to be localized only on the outer layer (top or bottom depending on the sign of $D$), causing multiple changes in the electronic magnetization. Irrespective of these details, the key conclusion is that the magnetoconductance hysteresis is observed only in even CrPS$_4$ multilayers, due to inversion symmetry caused by a finite displacement field.

**Doping -dependent hysteresis**

Finally, we discuss the evolution of the magnetoconductance hysteresis in bilayers as the carrier density is increased past the value at which the uniaxial magnetic anisotropy $K$ vanishes. If the sign of $K$ changes, as we expect for $n_e > 7\text{-}8 \times 10^{12}$ cm$^{-2}$, the magnetization in 2 L CrPS$_4$ reorients to point in-plane. States A and B then have the same energy at a finite magnetic field, and no switching –hence no magnetoconductance hysteresis– should be observed. This is what we find if we measure the magnetoconductance at a fixed $D$ ($D/\varepsilon_0 = -0.26$ V/nm in Fig. 5a) for increasing values of $n_e$: the hysteresis becomes less pronounced, the coercive field decreases, and vanishes around $n_e \approx 7\text{-}8 \times 10^{12}$ cm$^{-2}$. Fig. 5b and 5c summarize the evolution of the hysteretic part of the magnetoconductance ($\delta G_\uparrow - \delta G_\downarrow$). Fig. 5d shows that the coercive field extracted from this data scales with $n_e$ as the uniaxial magnetic anisotropy constant (see Fig. 2h). Consistently, we also see that the spin-flop transition – proportional to $\sqrt{K}$– shifts to lower magnetic fields, becomes less pronounced, and eventually vanishes. Similar trends are observed for 4 L CrPS$_4$ (see Supplementary Fig. 6).

**Conclusions**

The switchable magnetotransport hysteresis reported here shows directly the relation between spatial symmetry breaking –inversion symmetry in the present case– and the existence of spin-polarized bands in antiferromagnets. Controlling spatial symmetries employing field-effect transistors based on 2D antiferromagnetic semiconductors introduces new functionalities, possibly relevant for antiferromagnetic spintronics, such as the ability to switch on and off spin-dependent transport at will, in principle at high frequency. Furthermore, double-gated CrPS$_4$ bilayers provide a very promising platform to realize gate-induced conductors with near unity spin polarization[50], i.e., transistors in which not only the flow of charge but also the flow of spin is controlled. The evolution of the amplitude of hysteretic magnetoconductance with displacement field (Fig. 3d) strongly suggests that at a sufficiently large $D$ and low electron density, only one spin-polarized band is occupied, implying that in double-gate CrPS$_4$ bilayer transistors the dominant spin of the electrons responsible for charge transport can be controllably switched by gate. Conceiving feasible experiments to validate this conclusion –i.e., to directly measure the spin polarization of the accumulated electrons as a function of displacement field– is a key milestone for upcoming work.

Finally, when the mobility of charge carriers will be improved, we expect double gated transistors of CrPS$_4$ bilayers to give control over the sign of anomalous Hall effect. At zero perpendicular electric displacement field –in the absence of spin-polarization– no anomalous Hall effect should be observed. As the displacement field is turned on, the anomalous Hall effect should emerge with a sign determined by the sign of $D$, which fixes the sign of the spin polarization. Gate switchable spin-polarization and anomalous Hall effect have never been investigated earlier in bilayer antiferromagnets, and we expect that improving the material quality may allow exploring these phenomena and discovering others.

**Acknowledgments**

The authors gratefully acknowledge Alexandre Ferreira for technical support and useful discussions with Volodymyr Multian, Dmitry Lebedev, Nicolas Ubrig. A. F. M. gratefully acknowledges the Swiss National Science Foundation (Division II, project #200020_178891) and the EU Graphene Flagship project for support. M.G. acknowledges support from Ministero Italiano dell'Università e della Ricerca through the PNRR project ECS_00000033_ECOSISTER and the PRIN2022 project SECSY. K.W. and T.T. acknowledge support from the JSPS KAKENHI (Grant Numbers 21H05233 and 23H02052) and World Premier International Research Center Initiative (WPI), MEXT, Japan.

**Author Contributions**

F.Y., M.L. and A.F.M. initiated the work on atomically thin layers $CrPS_4$ field effect transistors. F.Y. and M.L. fabricated the devices and performed the transport measurements with the assistance of I.G.L, C.Y.C., X.L. and F.W.. M.G. performed the theoretical calculations. T.T and K.W grew and provided the h-BN crystals. F.Y., M.G., M.L., I.G.L and A.F.M. analyzed the data and wrote the manuscript with input from all authors. I.G.L. and A.F.M supervised the research.

**Competing Interests:** The authors declare no competing interests.

**Figure Legends**

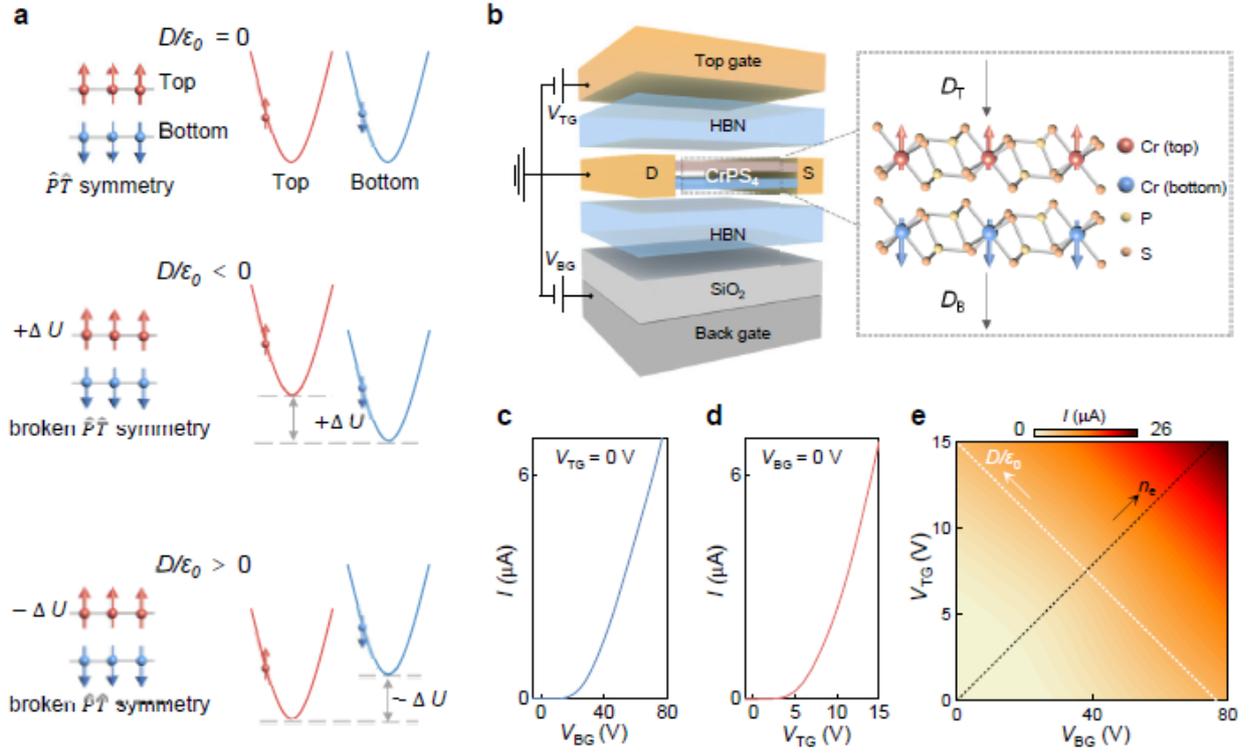

**Fig. 1: Switching spin polarized bands in A-type antiferromagnetic bilayers (2L). a**, Schematics of the low-energy band structure. An A-type antiferromagnetic bilayer hosts switchable spin-up (-down) bands, located in separate layers. The red (blue) spin-up (-down) bands represent the dispersion relation of states in the top (bottom) layer. At zero displacement field ($D/\varepsilon_0 = 0$, top panel), the potential difference $\Delta U$ between the layers vanishes, and the spin up and down bands are degenerate. At finite $\Delta U$ ($D/\varepsilon_0 \neq 0$, middle and bottom panel), the interlayer potential difference lowers the energy of states in one layer (hence with one spin polarity) relative to the other, thereby breaking inversion ($\hat{P}$) and space-time ($\hat{P}\hat{T}$) symmetry. Reversing the sign of $D/\varepsilon_0$ lowers the energy of states located in the other layer (bottom panel), with opposite spin (see also Supplementary Fig. 1). **b**, Schematic representation of a double-gated $CrPS_4$ bilayer transistor of the type employed in this work (see Methods for fabrication details). The zoom-in panel shows a side view of the $CrPS_4$ crystal structure (the red, blue, yellow and orange balls represent top-layer Cr, bottom-layer Cr, P and S atoms, respectively). The double gated geometry enables independent control of the electron density ($n_e$) and displacement field ($D/\varepsilon_0 = (D_T + D_B)/2\varepsilon_0$; $D_T$ ($D_B$) is the displacement field generated by the top (bottom) gate). **c**, Source (S)-drain (D) current $I$ measured in the transistor as a function of top-gate voltage $V_{TG}$ (transfer curve) at fixed back-gate voltage $V_{BG} = 0$. **d,** Transfer curve measured as a function of $V_{BG}$ for $V_{TG} = 0$. **e**, Color plot of $I$ as a function of $V_{BG}$ and $V_{TG}$. The black (white) dotted line corresponds to a constant density (displacement) profile. Data shown in this figure were measured at $V_{SD} = 2V$ and at 2 K.

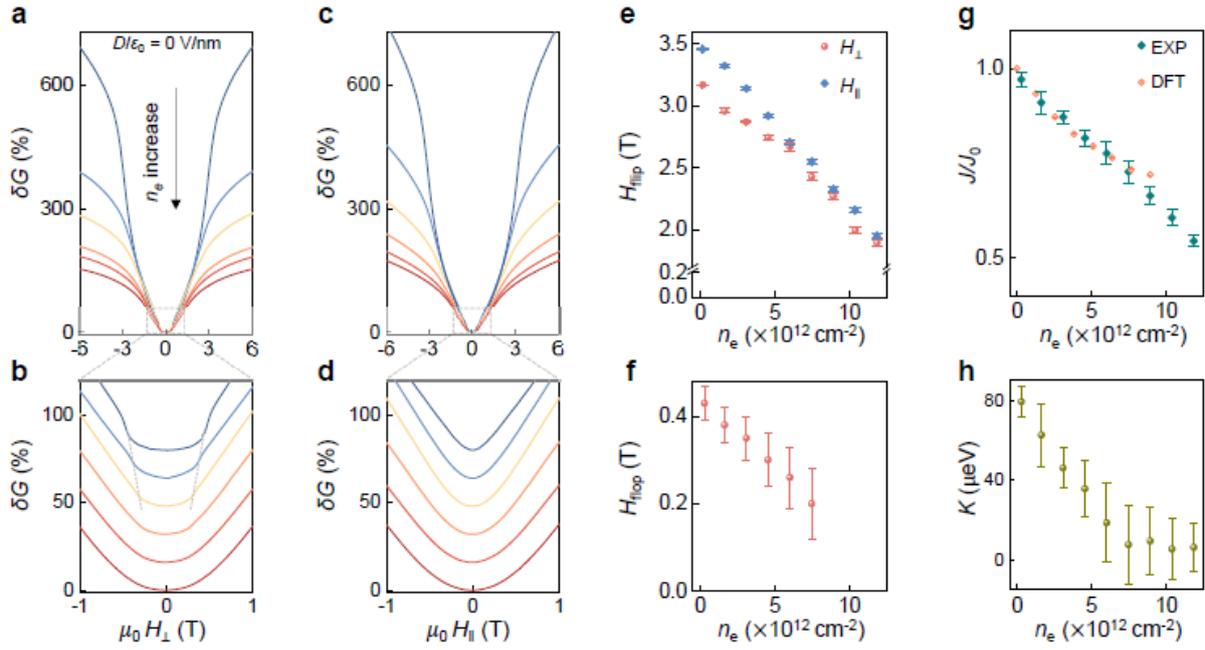

**Fig. 2: Doping-dependent magnetism in 2L CrPS$_4$ at zero displacement field. a**, Magnetoconductance $\delta G$ measured at 2 K and zero $D/\varepsilon_0$ as a function of out-of-plane magnetic field $H_\perp$, for $n_e$ varying from $1.75 \times 10^{11}$ to $1.2 \times 10^{13}$ cm$^{-2}$. The field at which $\delta G$ starts to flatten corresponds to the spin-flip field ($H_{\text{flip}}$), and can be precisely determined from the minimum in the second derivative of conductance ($G$) with respect to $H_\perp$ (see **c**). **b**, Low-field magnetoconductance, showing the effect of the spin-flop field ($H_{\text{flop}}$, indicated by dash grey line; $H_{\text{flop}}$ is determined by the position of the corresponding peak in $d^2G/dH^2$). **c**, $\delta G$ measured under the same conditions as in **a**, with field $H_\parallel$ applied parallel to the plane (as expected, the spin-flop transition is absent; see panel **d**). **e**, Evolution of $H_{\text{flip}}$ (obtained from the minimum in $d^2G/dH^2$; see Supplementary Fig. 2) for $H_\perp$ and $H_\parallel$ (red and blue dots, respectively), as a function of $n_e$. $H_{\text{flip}}$ is slightly smaller when the field is applied perpendicular to the plane (see **e**), because of the uniaxial magnetic anisotropy of CrPS$_4$ (see Supplementary Note 2). **f**, Evolution of $H_{\text{flop}}$ for $H_\perp$, as a function of $n_e$. **g**, $n_e$-dependent interlayer exchange energy ratio $J/J_0$ (cyan diamonds for experimental data, orange diamonds for density functional theory calculations). $J_0$ is the interlayer exchange energy at the lowest doping, $J_0^{\text{DFT}}$ ($n_e = 0$) = 1.1 meV for the DFT calculations, and $J_0^{\text{EXP}}$ ($n_e = 1.75 \times 10^{11}$ cm$^{-2}$) = 0.58 meV for the experimental values (a trend towards flattening seen in the DFT calculations at large $n_e$ remains to be understood). **h**, Uniaxial magnetic anisotropy $K$ extracted from $H_{\text{flip}}$ measured for both $H_\perp$ and $H_\parallel$ (see Supplementary Note 2 for details). The error bars in **e** and **f** are estimated from the width of the maximum/minimum in $d^2G/dH^2$ used to determine $H_{\text{flop}}$ and $H_{\text{flip}}$ (the width is taken at a 1% deviation from the maximum/minimum; see Supplementary Fig. 3 for an example). The error bars in **g** and **h** are calculated by propagating the errors shown in panel **e**.

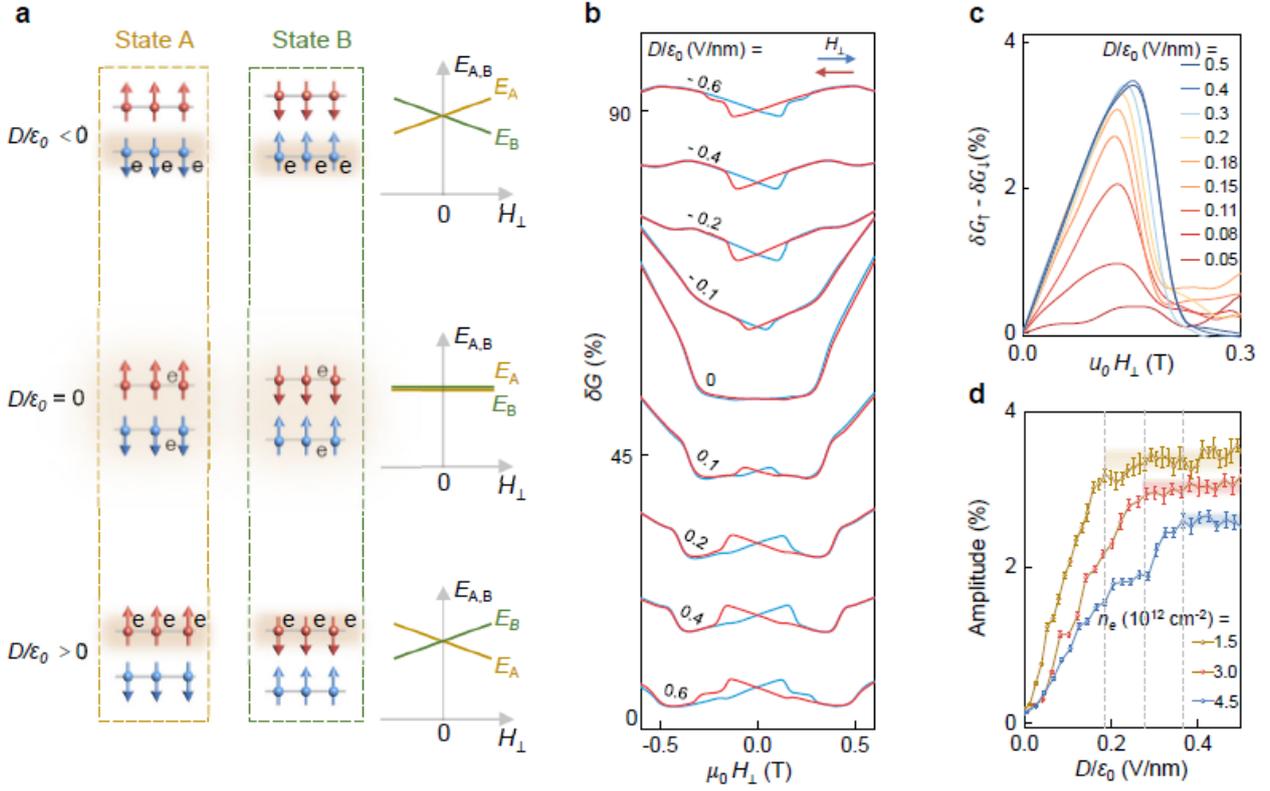

**Fig. 3: Detecting switchable spin polarized bands in 2 L CrPS$_4$. a,** At small magnetic field, 2L CrPS$_4$ can be in one of two states –with opposite magnetization in the top and bottom layers– labelled as A (left column) and B (right column). These states are energetically degenerate ($E_A = E_B$) in the absence of gate-induced electrons ($n_e = 0$) or when electrons equally occupy both layers ($D/\varepsilon_0 = 0$). The degeneracy is broken at finite $n_e$ and perpendicular electric field $D/\varepsilon_0$, in the presence of a perpendicular magnetic field. The state with lower energy is determined by the sign of the applied electric and magnetic fields, because the energy difference arises from the Zeeman energy of accumulated electrons that –a low $n_e$– occupy states in one of the two layers (and are therefore spin-polarized with spin pointing in opposite directions; see Fig. 1a). **b,** $\delta G$ measured at different $D/\varepsilon_0$ (see legend) at $n_e = 1.5 \times 10^{12}$ cm$^{-2}$. At $D/\varepsilon_0 = 0$, no hysteresis is observed due to the degeneracy of states A and B, as expected. The hysteresis that appears at finite $D/\varepsilon_0$ when sweeping $H_\perp$ up or down provides direct experimental evidence for the existence of the two states with different energies in a finite perpendicular magnetic field. **c,** Magnetoconductance hysteresis, corresponding to the difference between the magnetoconductance measured with field swept up ($\delta G_\uparrow$) or down ($\delta G_\downarrow$) for different values of $D/\varepsilon_0$ (see legend) and constant density $n_e = 1.5 \times 10^{12}$ cm$^{-2}$. The hysteresis amplitude first increases with $D/\varepsilon_0$, then saturates. **d,** Hysteresis amplitude as a function of $D/\varepsilon_0$ for different carrier densities. The saturation threshold of $D_{sat}/\varepsilon_0$ shifts to higher values as the carrier density increases, as indicated by the dashed lines and shaded regions. The error bars in **d** are estimated from the noise level of the curves in panel **c**.

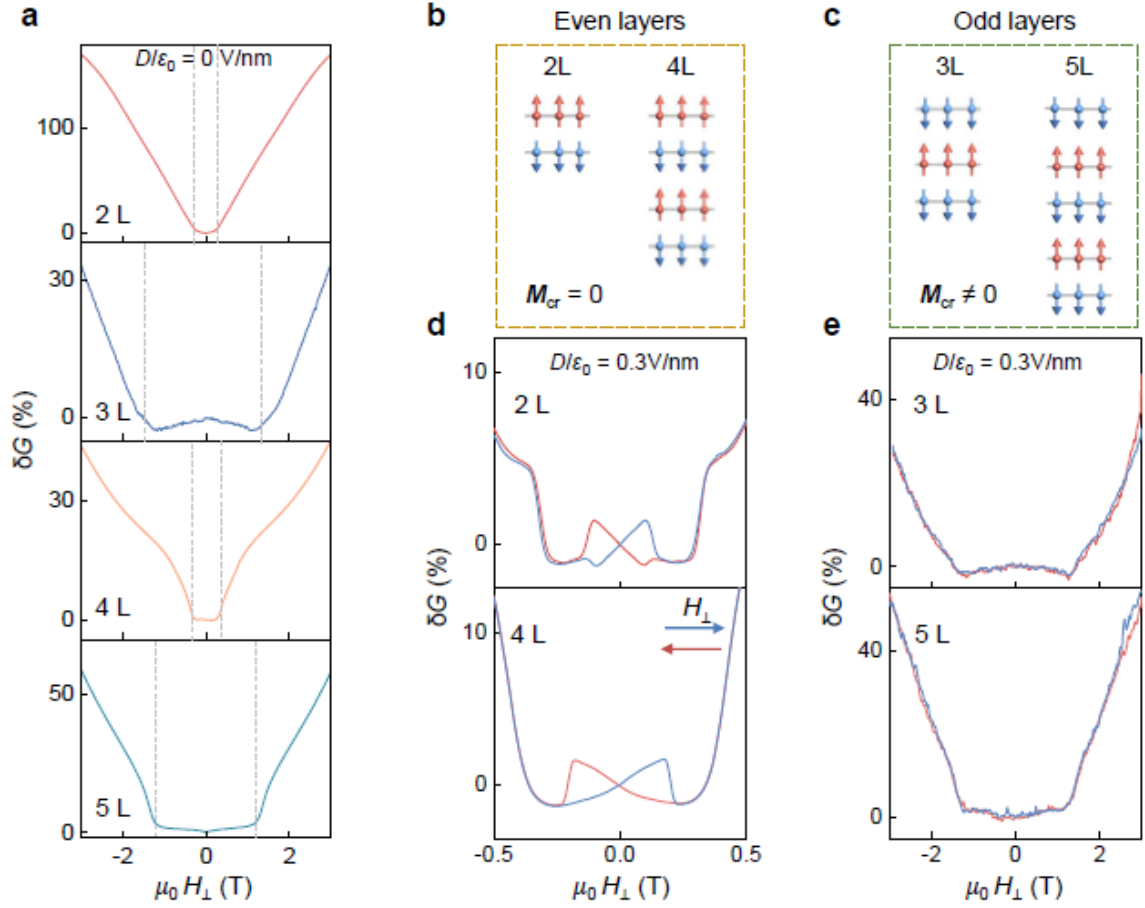

**Fig. 4: Odd-even effect of magnetoconductance hysteresis. a,** Magnetoconductance $\delta G(H)$ measured at 2 K and zero electric field ($D/\varepsilon_0 = 0$, $n_e \approx 4.5 \times 10^{12}$ cm$^{-2}$) on double-gated devices based on 2 L, 3 L, 4 L and 5 L CrPS$_4$ (from top to bottom, see legends). The spin-flop transition (indicated by the dashed lines) in odd layers occurs at 2-3 times larger field than in even layers (approximately 0.5 T), as expected for weakly anisotropic layered antiferromagnets[43, 49]. In the absence of an applied displacement field, no magnetoresistance hysteresis is observed irrespective of layer thickness. **b,** Schematic representation of the A-type magnetic order in even number of CrS$_4$ layers, where red (blue) arrows represent the magnetization of the Cr atoms in each layer. For even layers the total magnetization due to the Cr atoms vanishes. **c,** Same as in **b** but for odd number of CrS$_4$ layers. The total magnetization of Cr atoms is finite due to the presence of an unpaired layer. **d,** Magnetoconductance $\delta G(H)$ measured at 2 K on even layers (2 L and 4 L) at $D/\varepsilon_0 = 0.3$ V/nm and $n_e \approx 4.5 \times 10^{12}$ cm$^{-2}$, exhibiting a clear hysteresis when sweeping the perpendicular magnetic field ($H_\perp$). **e,** Magnetoconductance $\delta G$ ($H$, 2 K) measured on odd layers (3 L and 5 L) at $D/\varepsilon_0 = 0.3$ V/nm and $n_e \approx 4.5 \times 10^{12}$ cm$^{-2}$, exhibiting no hysteresis below spin-flop field even when an applied perpendicular displacement field is present.

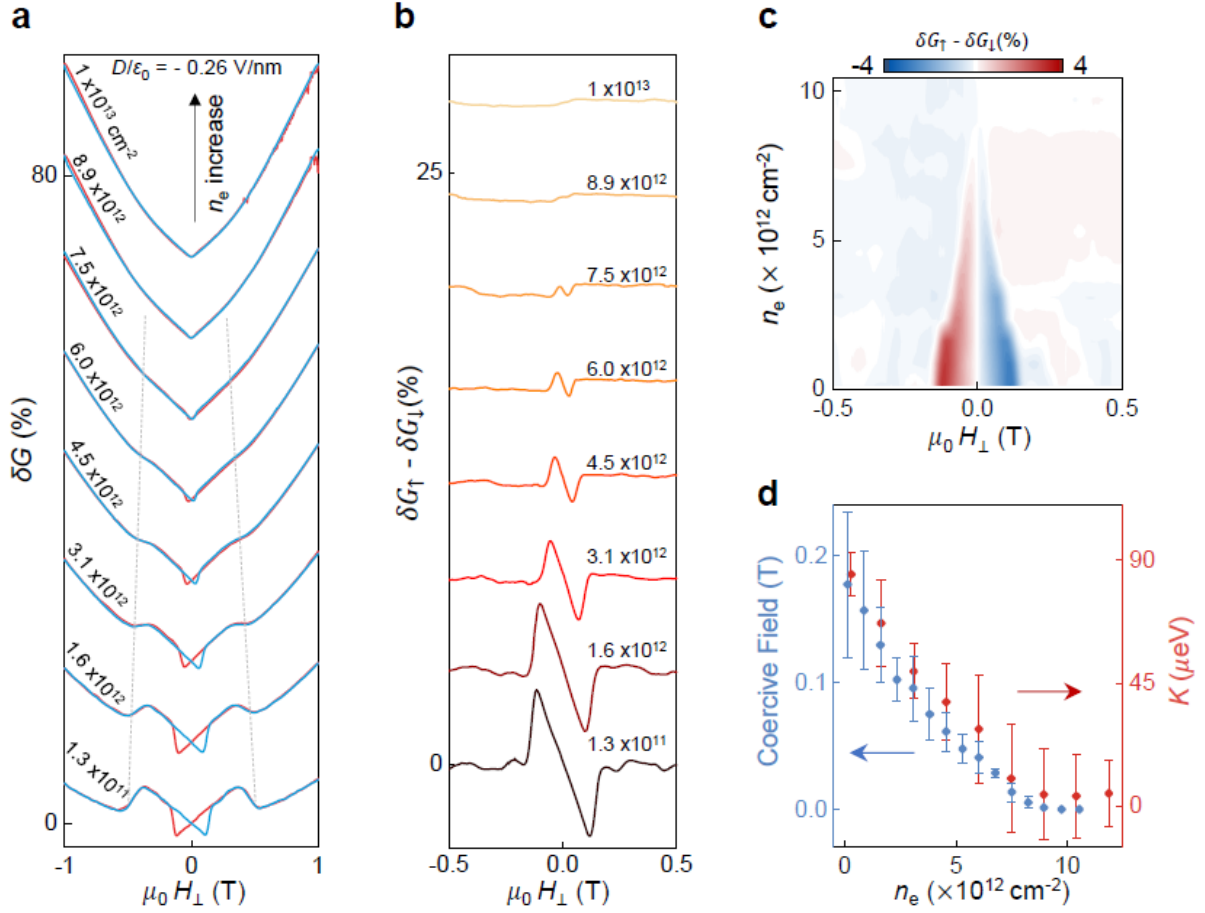

**Fig. 5: Dependence of magnetic state on electron density. a,** Low-field magnetoconductance $\delta G(H_\perp)$ measured for increasing values of accumulated electron density $n_e$ (see legend), at constant $D/\varepsilon_0 = -0.26$ V/nm (see Supplementary Fig.7 for analogous data at positive $D/\varepsilon_0$). The hysteresis caused by the finite magnetization of electrons occupying spin-polarized bands becomes less pronounced –and the magnetic field at which the hysteresis ends (i.e., the coercive field) becomes smaller– upon increasing $n_e$. Both quantities eventually vanish at approximately $7\text{-}8 \times 10^{12}$ cm$^{-2}$. The feature associated to the spin-flop transition (indicated by the grey dotted lines) also shifts to lower magnetic fields, becomes less pronounced, and eventually disappears on the same $n_e$ scale. **b,** Amplitude of the magnetoconductance hysteresis ($\delta G_\uparrow - \delta G_\downarrow$) obtained from **a** for different values of $n_e$, and **c,** color plot of $\delta G_\uparrow - \delta G_\downarrow$ as a function of $n_e$ and $H_\perp$, both showing that the hysteresis vanishes at $n_e \approx 7\text{-}8 \times 10^{12}$ cm$^{-2}$. **d,** The carrier density evolution of the coercive field extracted from **b** and **c**, and of the uniaxial magnetic anisotropy $K$ extracted from the analysis of the spin-flip field in parallel and perpendicular magnetic field (see Fig. 2e), show that the two quantities are proportional. The error bars for the coercive field correspond to the width of the jump shown in panel **b**. For $K$, the error bars in **d** are reported from Fig.2e.

39 Wu, F. et al. Magnetism-induced band-edge shift as the mechanism for magnetoconductance in CrPS$_4$ transistors. *Nano Lett.* **23,** 8140-8145 (2023).

40 Wang, Z. et al. Very large tunneling magnetoresistance in layered magnetic semiconductor CrI$_3$. *Nat. Commun.* **9,** 2516 (2018).

41 Li, J., Gutierrez-Lezama, I., Morpurgo, A. F. Magneto-transport study in 2D magnetic semiconductor multi-terminal FET [Zenodo]. *https://doi.org/10.5281/zenodo.12702065* (2024).

42 Chang, J.-F. et al. Hall-Effect measurements probing the degree of charge-carrier delocalization in solution-processed crystalline molecular semiconductors. *Phys. Rev. Lett.* **107,** 066601 (2011).

43 Wang, Z. et al. Determining the phase diagram of atomically thin layered antiferromagnet CrCl$_3$. *Nat. Nanotechnol.* **14,** 1116-1122 (2019).

44 Yao, F. et al. Multiple antiferromagnetic phases and magnetic anisotropy in exfoliated CrBr$_3$ multilayers. *Nat. Commun.* **14,** 4969 (2023).

45 Tang, M. et al. Continuous manipulation of magnetic anisotropy in a van der Waals ferromagnet via electrical gating. *Nat. Electron.* **6,** 28-36 (2023).

46 Clark, A. E., Callen, E. Néel ferrimagnets in large magnetic fields. *J. Appl. Phys.* **39,** 5972-5982 (1968).

47 Coey, J. M., *Magnetism and magnetic materials* (Cambridge university press, 2010).

48 Zhuang, H. L., Zhou, J. Density functional theory study of bulk and single-layer magnetic semiconductor CrPS$_4$. *Phys. Rev. B.* **94,** 195307 (2016).

49 Ye, C. et al. Layer-dependent interlayer antiferromagnetic spin reorientation in air-stable semiconductor CrSBr. *ACS nano* **16,** 11876-11883 (2022).

50 Son, Y.-W., Cohen, M. L., Louie, S. G. Half-metallic graphene nanoribbons. *Nature* **444,** 347-349 (2006).

## Methods

**Device fabrication**

The h-BN/CrPS$_4$/graphite (Gr)/h-BN heterostructures were assembled using a dry pick-up and transfer technique with PDMS-PC stamps in an N$_2$-filled glove box (H$_2$O < 0.1 ppm, O$_2$ < 0.1 ppm). CrPS$_4$ multilayers were obtained via micromechanical exfoliation inside the glove box from bulk crystals purchased from HQ Graphene. The chemical composition and stoichiometry of CrPS$_4$ bulk crystals were verified using energy-dispersive X-ray spectroscopy (EDS) in a scanning electron microscope, with elemental mapping showing a uniform distribution and an atomic ratio of Cr:P:S as 16.92:16.60:66.48, consistent with the expected 1:1:4 stoichiometry[38]. Graphene and h-BN flakes were prepared by mechanical exfoliation onto SiO$_2$/Si substrates. The CrPS$_4$ crystals were encapsulated with top and bottom h-BN layers. Separate graphite stripes acted as source-drain electrodes, and were connected to metallic pads via edge contacts located far from the CrPS$_4$ crystal. The edge contacts and metallic pads were fabricated using electron beam lithography, reactive-ion etching, electron-beam evaporation of 10 nm Cr followed by 50 nm Au, and a lift-off process. Cr/Au contact electrodes were also deposited on the top h-BN to form the top gate. As bottom gate electrode we used the highly doped Si substrate, with the SiO$_2$ layer serving as gate dielectric together with the bottom h-BN crystal. In total, we fabricated nine devices using thin CrPS$_4$ layers. We fabricated one device each for 2 L, 3 L and 4 L in a single-gate configuration, with only the bilayer and four-layer devices exhibiting hysteresis. To study the independent effects of the electric field and doping, we fabricated two 2 L devices, two 4 L devices, one 3 L device, and one 5 L device in a double-gate configuration. The thickness of CrPS$_4$ was determined using optical contrast and Raman spectroscopy (see Extended Data Fig. 2 for details).

**Transport measurement**

Low-noise homemade electronics in combination with commercial electronics was employed to bias the top and bottom gate electrodes, to apply source-drain voltage, and to measure the current. Top-gate voltage ($V_{TG}$) and bottom-gate voltage ($V_{BG}$) were swept to adjust the doping density ($n_e = [C_t (V_{TG} - V_{TGTH}) + C_b (V_{BG} - V_{BGTH})]/e$) and the electric displacement field ($D/\varepsilon_0 = [(V_{TG} - V_{TGTH})/d_t - (V_{BG} - V_{BGTH})/d_b]/2$). $C_t$ and $C_b$ are top- and bottom-gate capacitance per unit area; $V_{TGTH}$ and $V_{BGTH}$ are threshold voltages when sweeping the back gate and top gate respectively; $d_t$ is the thickness of the top hBN; $d_b$ is the combined thickness of bottom h-BN and SiO$_2$ layers. Low-temperature transport measurements were conducted in an Oxford Instruments cryostat equipped with a superconducting magnet and a $^3$He insert.

**Antiferromagnetic two-site model**

To model the energetics of the bilayer, we assume that the ferromagnetic intralayer exchange coupling is so strong (compared to the weak antiferromagnetic interlayer exchange coupling and the external magnetic field) that each layer can be considered as a single unit with uniform magnetization. Each layer thus behaves as a macroscopic spin that is coupled antiferromagnetically to its neighbor, so that the average magnetic energy per unit cell can be written as[43]:

$$E = J\mathbf{M}_1 \cdot \mathbf{M}_2 / M_s^2 - K/2 \, (M_{1z}/M_s)^2 - K/2 \, (M_{2z}/M_s)^2 - \mu_0 \mathbf{H} \cdot (\mathbf{M}_1 + \mathbf{M}_2),$$

where $J$ is the antiferromagnetic interlayer exchange coupling, $K$ is the magnetic anisotropy energy favoring out-of-plane orientation, $\mathbf{M_1}$ and $\mathbf{M_2}$ are the magnetization vectors of the two layers, and $\mathbf{H}$ is the applied magnetic field. Here $M_s = 2g\mu_B S$ is the saturation magnetization (per unit cell) for a single layer, which can be easily computed from the nominal valence state of Cr atoms in CrPS$_4$ (corresponding to $S = 3/2$).

**Density functional theory calculations**

The total energy of the bilayer in the ferromagnetic and antiferromagnetic configurations have been computed from first principles using density functional theory as implemented in Quantum ESPRESSO distribution[51, 52].

Atomic positions for the bilayer have been extracted from the bulk relaxed crystal structure[38]. We have verified that, relaxing the atomic positions, the interlayer distance in the bilayer is only marginally changed with respect to the bulk material. The total energy is obtained adopting the Perdew-Burke-Ernzerhof (PBE) exchange-correlation functional[53] with pseudopotentials taken from the Standard Solid-State Pseudopotential (SSSP) accuracy library (v1.0)[54] (cutoffs of 40~Ry and 320~Ry for wave functions and density). Hubbard corrections have not been included in the calculations. The interlayer exchange energy (per unit cell) is then evaluated as half the energy difference between the ferromagnetic and antiferromagnetic configurations, $J = (E_{FM} - E_{AFM})/2$. The 2D nature of the system is taken into account by using a Coulomb cutoff while the effect of doping is simulated using a double-gate field-effect setup[55]. To sample the small Fermi surface at finite doping, a dense $24 \times 24 \times 1$ Monkhorst-Pack grid over the Brillouin zone is adopted, with a Gaussian smearing of 6.4 meV.

**Data availability**

All relevant data are available from the corresponding authors upon request.

**Code availability**

All codes adopted for DFT calculation are available from the corresponding authors upon request.

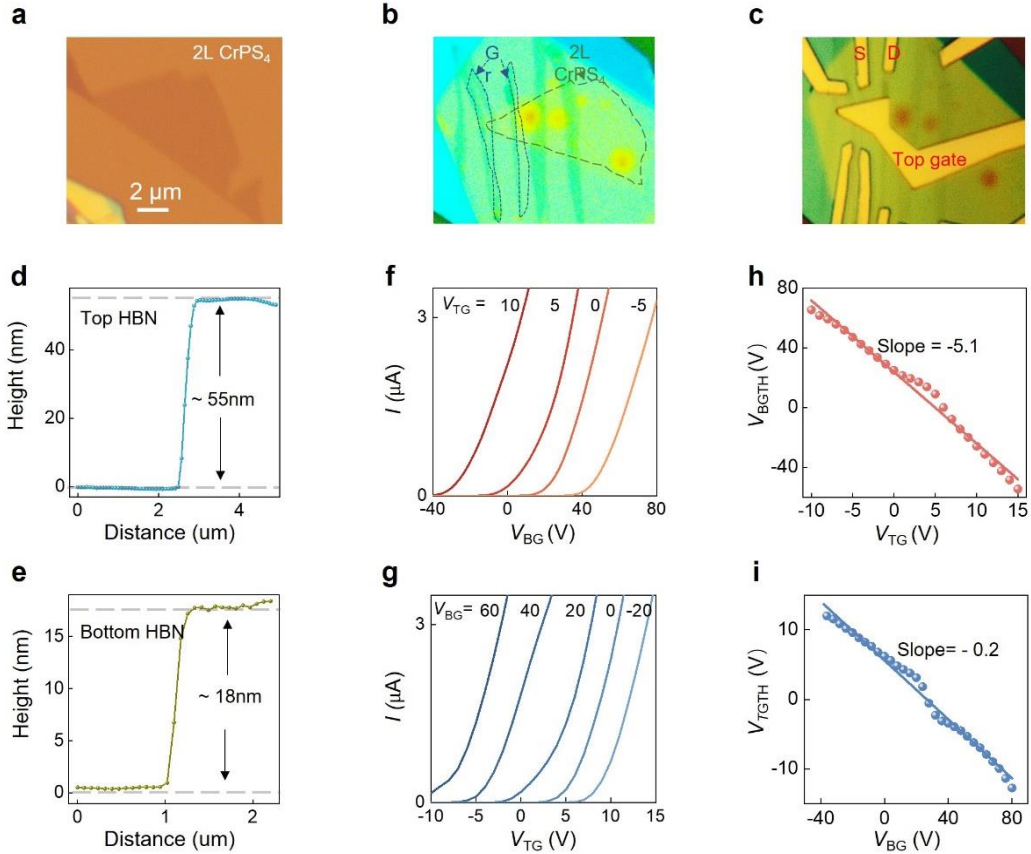

**Extended Data Fig. 1**. **Double-gated bilayer (2L) CrPS$_4$ field effect transistor (FETs). a-c,** Optical microscope images of the 2L CrPS$_4$ FET whose data is shown in the main text, at different stages of fabrication. **a,** Exfoliated 2L CrPS$_4$ crystal on top of a 285 nm SiO$_2$/Si substrate. **b,** h-BN/ CrPS$_4$ /graphite (Gr)/h-BN heterostructure assembled via a dry-pick up and transfer technique within the protective environment of a glove box. The graphite stripes serve as electrodes for the transistor, and the top and bottom h-BN crystals encapsulate the 2L CrPS$_4$ to protect it from the environment during subsequent fabrications steps outside the glove box. **c,** Completed double-gated FET. Electron-beam lithography and evaporation are used to fabricate the Cr/Au metal layers serving as contacts to the Gr electrodes and as top gate electrode (top h-BN serves as top gate dielectric). The bottom gate electrode is the highly doped Si substrate and the bottom gate dielectric is composed by the 285 nm SiO$_2$ layer together with the bottom h-BN crystal. **d-e,** Atomic force microscope height profile of the top and bottom h-BN crystals. **f, g,** Transfer curves measured at 2 K as a function of bottom (top) gate voltage $V_{BG}$ ($V_{TG}$), while keeping the top (bottom) gate at a constant potential (at the values indicated in the figures). **h, i,** Back gate and top gate and threshold voltage, $V_{BGTH}$ and $V_{TGTH}$, plotted as a function of the potential applied to the opposite gate. The slope of $V_{BGTH}$ vs $V_{TG}$ in **f** and of $V_{TGTH}$ vs $V_{BG}$ in **g** are consistent with the corresponding ratios between the top and bottom gate capacitances. The threshold voltage values in **h** (**i**) are extracted by extrapolating the linear regime of the transfer curves shown in **f** (**g**) to zero current.

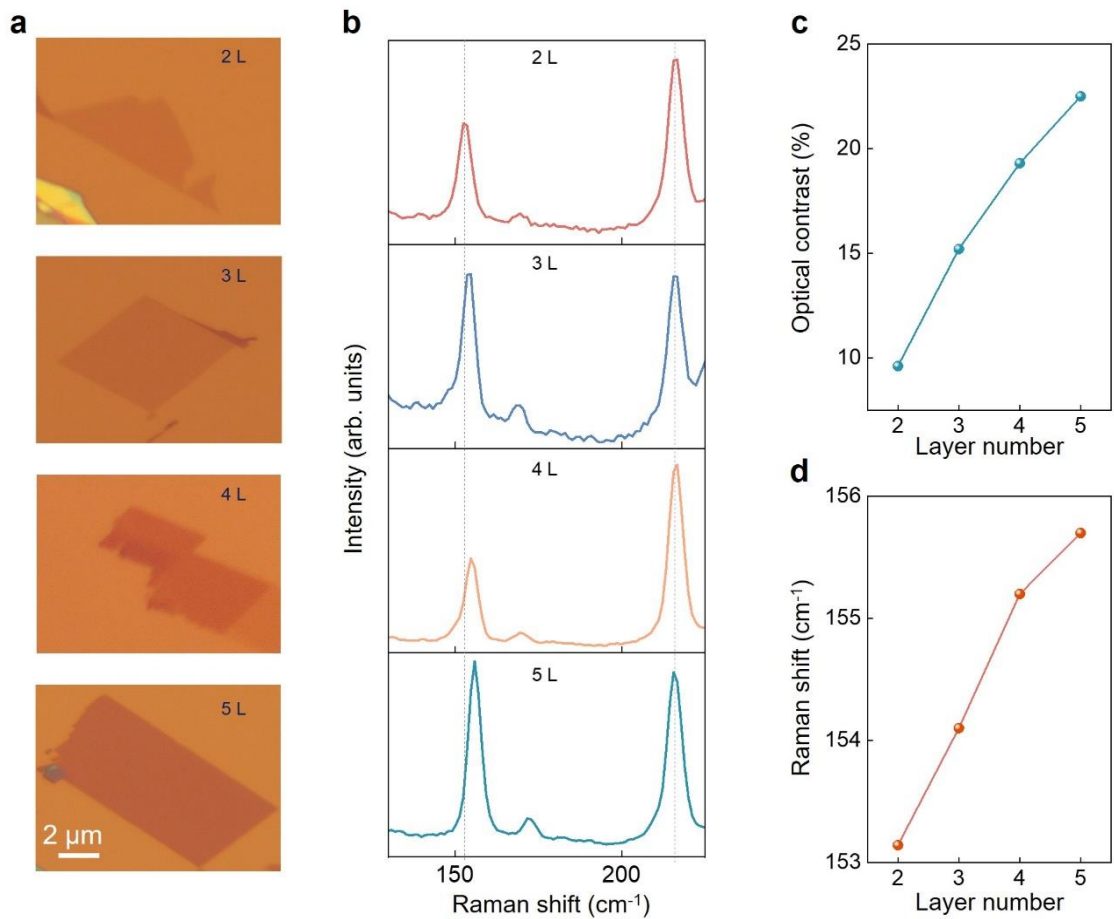

**Extended Data Fig. 2**. **Thickness characterization of atomically thin CrPS₄ multilayers. a,** Optical micrographs of CrPS₄ multilayers with thicknesses ranging from 2L to 5L. **b,** Raman spectra of CrPS₄ multilayers, showing a redshift in the peak near 150 cm⁻¹ as thickness increases, consistent with previously reported thickness-dependent shifts[34, 37]. **c,** Extracted optical contrast value of CrPS₄ multilayers with thicknesses from 2L to 5L. **d,** Extracted peak positions of the Raman mode around 150 cm⁻¹ for multilayers with thicknesses from 2L to 5L. These results demonstrate that the layer number of thin CrPS₄ can be identified through the redshift in the Raman peak and corresponding changes in optical contrast.

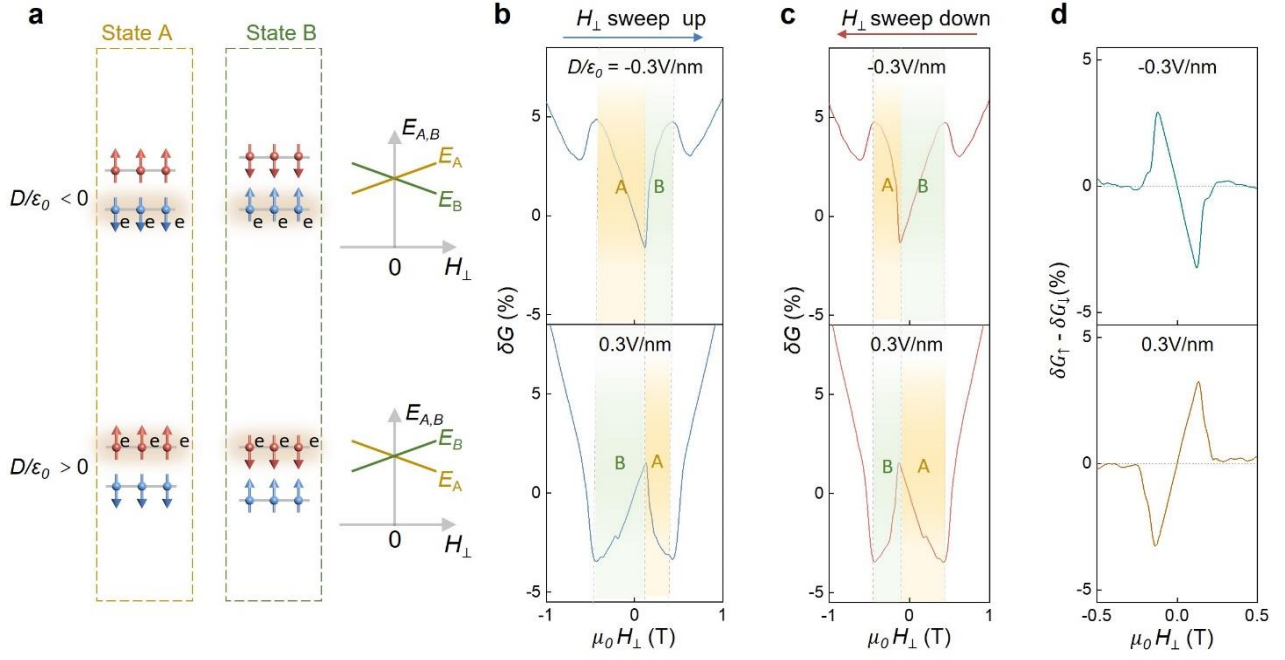

**Extended Data Fig. 3**. **Identification of states A and B in terms of a measurable quantity. a,** At small magnetic fields, 2L CrPS$_4$ can exist in two magnetic states, A and B, with opposite top and bottom layer magnetization. The spin polarization of gate-accumulated electrons (represented by the shaded region) follows the Cr atom magnetization, so that at finite $D/\varepsilon_0$ states A and B have different energies in the presence of a perpendicular magnetic field (Fig. 3). Here we show that the sign of $d(\delta G)/dH$ can be used to discriminate experimentally between the two states. **b-c,** Low-field magnetoconductance ($\delta G$) measured at 2 K at a constant electron density ($1.5 \times 10^{12}$ cm$^{-2}$) for opposite displacement fields, with the magnetic field ($H_\perp$) swept up (**b**) and down (**c**). There is a univocal correspondence between the sign of the derivative of $\delta G$ with respect to $H$ (i.e., $d(\delta G)/dH$) and whether the system is in state A or B. We label as "A", the state when $d(\delta G)/dH < 0$ (yellow shaded areas in panels **b** and **c**), and as "B" the state when $d(\delta G)/dH > 0$ (green shaded areas in panels **b** and **c**). Upon sweeping the magnetic field in different directions, and upon reversing the displacement field, the transitions between state A and B follow precisely the scheme outlined in panel **a**. For instance, in panel **a** for negative $D$, the bilayer undergoes a transition from state A to B when sweeping the magnetic field up, whereas for positive $D$ the transition is from B to A. Panel **b** shows that this is precisely what we conclude by looking at the sign of $d(\delta G)/dH$. **d,** Magnetoconductance hysteresis ($\delta G_\uparrow - \delta G_\downarrow$) for opposite values of $D/\varepsilon_0$ (see legend). Based on the discussed logic, the opposite signs of the hysteresis under reversed displacement fields can be attributed to the reversed order of state switching.

21 / 22

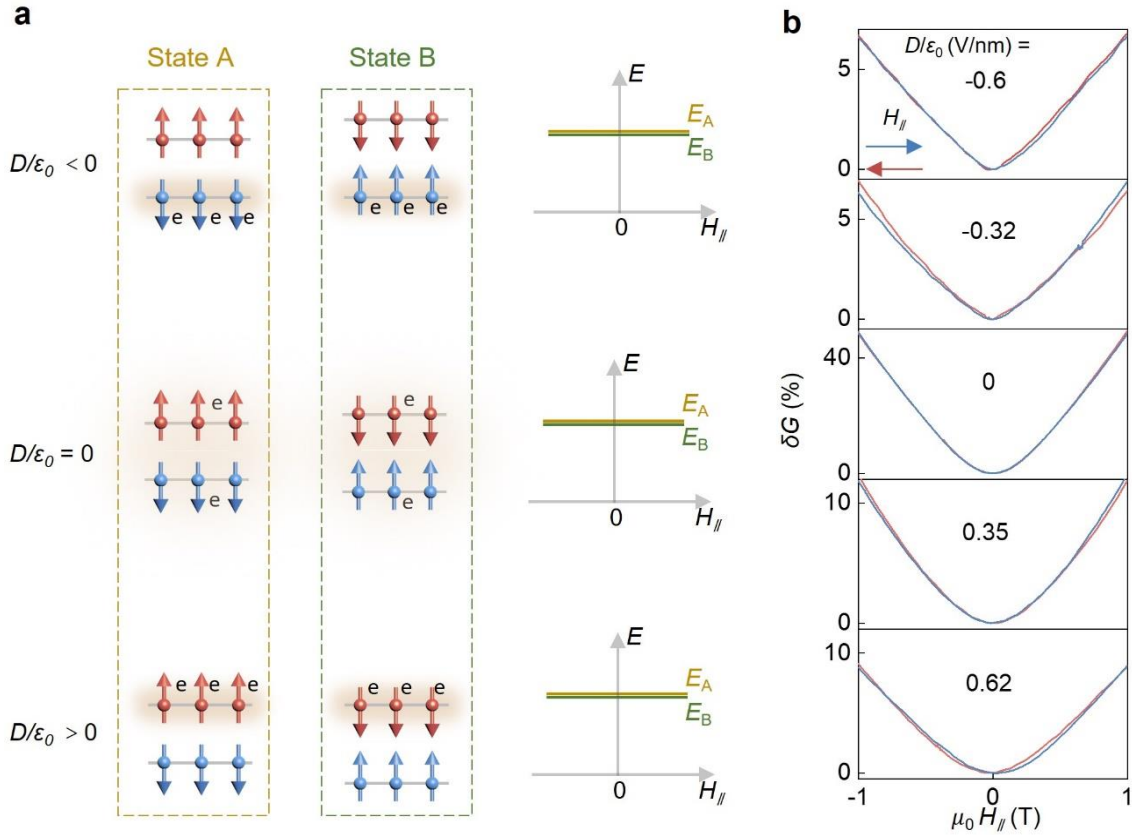

**Extended Data Fig. 4. Absence of hysteresis in the magnetoconductance under in-plane magnetic field. a,** As discussed in the main text, at small magnetic fields 2L CrPS$_4$ can be in one of two magnetic states, labelled A (left column) and B (middle column), where the magnetization of the top and bottom layers is opposite. The spin polarization of the electrons accumulated by the gate voltage (represented by the orange shadow) in each layer is determined by the magnetization of the Cr atoms in the same layer. When the magnetic field is applied perpendicular to the layers (as discussed in Fig. 3 of the main text), state A and B have different energy (at finite $D/\varepsilon_0$). However, when the magnetic field is applied parallel to the layers, the Zeeman energy of the electrons vanishes and state A and B always have the same energy. As a result, no hysteresis in the magnetoconductance is expected. **b,** Indeed $\delta G$ measured at $T = 2$ K for different displacement fields (from top to bottom: -0.6V/nm, -0.32V/nm, 0V/nm, 0.35V/nm, 0.62V/nm) is independent of the direction in which the magnetic field is swept (the red (blue) trace represents the magnetoconductance measured as the field is swept up (down)), i.e., no hysteresis is observed experimentally when the magnetic field is applied parallel to the plane, irrespective of the applied displacement field.